\documentclass[10pt,english,journal]{IEEEtran}
\usepackage[T1]{fontenc}
\usepackage[latin9]{inputenc}
\usepackage{geometry}
\usepackage{amsmath}
\usepackage{xpatch}
\usepackage{amssymb}
\usepackage{esint}
\usepackage{mathtools}
\usepackage{amsthm}
\usepackage{nicefrac}
\usepackage{bigints}
\usepackage{mathtools}
\usepackage{blkarray, bigstrut}
\usepackage{physics}
\usepackage{calligra}
\usepackage{graphicx}

\usepackage{dsfont}
\usepackage{array,ragged2e}
\usepackage{enumitem}
\usepackage{etoolbox}
\usepackage{babel}
\usepackage[nopar]{lipsum}
\usepackage{psfrag}
\usepackage[ruled,norelsize]{algorithm2e}
\usepackage{algorithmic}
\usepackage{algorithm2e}
\usepackage{balance}
\usepackage{float}
\usepackage{subcaption}

\makeatletter
\newcommand{\removelatexerror}{\let\@latex@error\@gobble}
\makeatother

\newtheoremstyle{plain}
  {\topsep}   
  {\topsep}   
  {\itshape}  
  {0pt}       
  {\bfseries} 
  {.}         
  {5pt plus 1pt minus 1pt} 
  {\thmname{#1}\thmnumber{ #2} \textnormal{(\thmnote{#3})}} 

\xpatchcmd{\proof}{\hskip\labelsep}{\hskip5\labelsep}{}{}  
\makeatletter
\xpatchcmd{\proof}{\@addpunct{.}}{\@addpunct{:}}{}{}
\makeatother

\renewcommand\[{\begin{equation}}
\renewcommand\]{\end{equation}} 
\usepackage{listings}
\usepackage{fancyvrb}
\usepackage{framed}

\usepackage{courier}
\usepackage[usenames,dvipsnames,table]{xcolor}

\definecolor{dkgreen}{rgb}{0,0.3,0}
\definecolor{gray}{rgb}{0.5,0.5,0.5}

\makeatletter
\newcommand*{\rom}[1]{\expandafter\@slowromancap\romannumeral #1@}
\makeatother

\begin{document}
\title{Continuous Multi-objective Zero-touch Network Slicing via Twin Delayed DDPG and OpenAI Gym}
\author{Farhad Rezazadeh$^1$, Hatim Chergui$^1$, Luis Alonso$^2$, and Christos Verikoukis$^1$\\
{\normalsize{} $^1$ Telecommunications Technological Center of Catalonia (CTTC), Barcelona, Spain\\ $^2$ Technical University of Catalonia (UPC), Barcelona, Spain}\\
{\normalsize{}Contact Emails: \texttt{farhad.rezazadeh, hatim.chergui, cveri@cttc.es, luisg@tsc.upc.edu}}}
\maketitle
\thispagestyle{empty}
\begin{abstract}
Artificial intelligence (AI)-driven zero-touch network slicing (NS) is a new paradigm enabling the automation of resource management and orchestration (MANO) in multi-tenant beyond 5G (B5G) networks. In this paper, we tackle the problem of cloud-RAN (C-RAN) joint slice admission control and resource allocation by first formulating it as a Markov decision process (MDP). We then invoke an advanced continuous deep reinforcement learning (DRL) method called twin delayed deep deterministic policy gradient (TD3) to solve it. In this intent, we introduce a multi-objective approach to make the central unit (CU) learn how to re-configure computing resources autonomously while minimizing latency, energy consumption and virtual network function (VNF) instantiation cost for each slice. Moreover, we build a complete 5G C-RAN network slicing environment using OpenAI Gym toolkit where, thanks to its standardized interface, it can be easily tested with different DRL schemes. Finally, we present extensive experimental results to showcase the gain of TD3 as well as the adopted multi-objective strategy in terms of achieved slice admission success rate, latency, energy saving and CPU utilization.
\end{abstract}


\section{Introduction}
\IEEEPARstart{N}{etwork} slicing is a key feature in 5G networks. It enables to run fully or partly isolated logical networks---or tenants---on the same physical network, offering thereby a concrete resource multiplexing gain between slice instances. In this intent, network softwarization and virtualization technologies such as software defined networking (SDN) and network functions virtualization (NFV) provide the necessary programmability and flexibility to operate NS by dynamically creating, scaling and terminating chained virtual network functions (VNFs). Multi-access Edge Computing (MEC) \cite{MEC} is also an important component that---co-located with C-RAN---can bring the high performance computing resources at the edge, paving the way to accommodate low-latency slices. Having said that, zero-touch and fully automated operations and management have become quintessential to harness the potential gain of dynamic resource allocation in SDN/NFV-enabled NS. Besides ETSI's architecture standardization efforts \cite{ZSM}, many algorithms have been presented in the literature to enable the automation of B5G networks as detailed in the sequel.
\subsection{Related Work}
In \cite{PAT}, the authors have studied autonomous MANO of VNFs, where the central unit learns to re-configure resources, deploy new VNF instances or offloaded to a central cloud. They have proposed a DRL-based solution dubbed \emph{parameterized action twin} (PAT) Deep Deterministic Policy Gradient (DDPG) which leverages the actor-critic method to learn to provision network resources to the VNFs in an online manner, given the current network state and the requirements of the deployed VNFs. The proposed solution outperforms all benchmark DRL schemes as well as heuristic greedy allocation in a variety of network scenarios. In \cite{DDDS}, the authors have developed a data-driven resource scheduling based on DRL for dynamic resource scheduling in networks slicing. They have solved the slicing resource management challenge in an asymmetric information scenario without using the user-related data, due to the model-free and dynamic online learning features. Correspondingly, \cite{DRDR} has proposed a dynamic resource reservation and DRL-based autonomous virtual resource slicing framework for the next generation radio access network. At light load, autonomous radio resource management of the deep $Q$-network (DQN) algorithm has achieved 100$\%$ satisfaction and up to about $80\%$ saturation which is the best compared with other benchmarks. Li \emph{et al.} have studied the application of DRL in some typical resource management scenarios of NS. Their results have shown that compared with the demand prediction-based and some other intuitive solutions, DRL could implicitly incorporate more deep relationship between demand and supply in resource-constrained scenarios \cite{RMIS}. In  \cite{vrAIn}, the authors have proposed vrAIn as a dynamic resource controller for virtualization of RANs (vRAN) based on DRL where vRAN dynamically learns the optimal allocation of computing and radio resources. The proposed solution meets the desired performance targets while minimizing CPU usage and gracefully adapts to shortages of computing resources.
\subsection{Contributions}
In this paper, we present the following contributions:
\begin{itemize}
    \item Given that DDPG algorithm is a limiting case of stochastic policy gradient in actor-critic approaches used for solving continuous tasks, this work adopts and fine-tune an alternative way of updating the actor (policy) in DDPG algorithm to speed-up convergence and fulfill a stable and robust learning process \cite{HPGDRL} based on TD3 method \cite{TD3}.
    \item We introduce a multi-objective approach in the NS environment to maximize cumulative rewards while minimizing network costs.
    \item We develop a complete 5G NS environment based on OpenAI Gym to ensure reproducible comparison of DRL algorithms.
\vspace{-0.1mm}
\end{itemize}

\section{System Model}
As depicted in Figure 1, we consider a C-RAN architecture according to 3GPP CU-DU functional split. The underlying $N$ single-antenna small-cells ($n=1,\ldots,N$) are connected to a virtual baseband unit (i.e., CUs) pool that runs as a set of VNFs. A total number of $J$ VNFs ($j=1,\ldots,J$) can be deployed on top of the C-RAN datacenter endowed with $I$ active central processing units (CPUs), where each processor $i$ ($i=1,\ldots,I$) has a computing capability of $P_i$ million operations per time slot (MOPTS) \cite{FOG}. At each time step $t$, $M$ UEs ($m=1,\ldots,M$) can connect to the $N$ small-cells according to the maximum received power criteria. Each UE $m$ requests a slice and starts its activity, wherein the packet arrival to the CU VNF follows a Poisson distribution with mean rate $\lambda_m^{(t)}$. In this case, let $\Omega = \sum_{m=1}^{M} {\lambda}_{m}^{(t)}$. The mean arrival data rate of all UEs to the CU VNFs is $\Omega/j$, where $j$ is the number of active VNFs.\\
\textbf{\textit{Computation cost}}$({K}_{Net}^{(t)})-$ The baseband processing procedure at a VNF consists of coding, Fast Fourier Transform (FFT) and modulation. The corresponding computing resources follow that in \cite{HMCC}, and is given by:
\vspace{-3mm}
\begin{equation}
   {K}_{Net}^{(t)}=\sum_{m=1}^{M}\left[\theta  \log_2(1+\delta_m) \right]+MK_0,
\end{equation}
Where $\theta$ is an experimental parameter, $\delta_m$ denotes the signal-to-interference-plus-noise ratio (SINR) of UE $m$ and $K_0$ includes computing resources for FFT function that according to \cite{AFFP} imposes a constant base processing load on the system. Based on the experimental results of \cite{PAOR} we assume that, in each cell $n$, the computing resource requirements for coding, modulation and FFT are 50\%,10\% and 40\%, respectively.
\begin{figure}[ht!]
\centering
\includegraphics[scale=0.4]{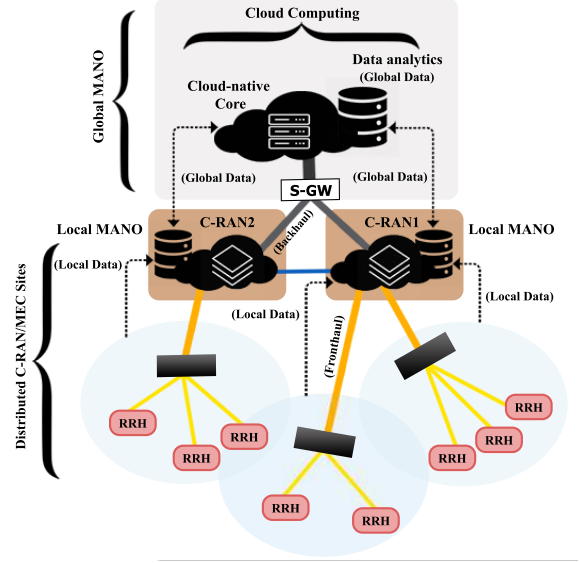}
\caption{Proposed distributed and hierarchical architecture.}
\end{figure}
Moreover, we assume that VNFs have first in first out (FIFO) queues where $\mu^*$ is mean service rate for cloud processing and $r_m$ for wireless transmission rate which satisfies according to $r_m = B_m\log(1+\delta_m)$, where $B_m$ is wireless transmission bandwidth for UE $m$. In this respect, we further suppose that cloud processing and wireless transmission queues follow an exponential distribution with mean $\frac{1}{\mu^*}$ and $\frac{1}{r_m}$  respectively \cite{SCMC}. Next, we explain each cost functions of this paper.\\
$\textbf{\textit{Latency}} (\mathcal{L}_{Net}^{(t)})-$ According to queuing theory \cite{QMFA}, the mean processing delay at time step $t$ is $\mathcal{L}_{proc}^{(t)} = \frac{j}{j{\mu^*}-\Omega}$ and transmission latency in wireless transmission queue is $\mathcal{L}_{trans}^{(t)} = \frac{1}{{r_m}-{\lambda}_{m}^{(t)}}$ so we have:
\begin{equation}
   \mathcal{L}_{Net}^{(t)} = j\mathcal{L}_{d}^{(t)} + \sum_{m=1}^{M} \left[ \frac{j}{j{\mu^*}-\Omega}+ \frac{1}{{r_m}-{\lambda}_{m}^{(t)}}\right]
\end{equation}
where $\mathcal{L}_{d}^{(t)}$ denotes latency for creating, booting up and loading new VNFs and $j$ denotes the total number of active VNFs to be deployed. We suppose  $\mathcal{L}_{Net}^{(t)}< \eta_{m}^{(t)}$ where $\eta_{m}^{(t)}$ is a predefined maximum network delay for UEs which can be viewd as a quality of service (QoS) requirement \cite{SCMC}.\\
$\textbf{\textit{Energy}} (\mathcal{E}_{Net}^{(t)}) -$ The energy consumption incurred by the VNF instantiation, running processors and wireless transmission power where $\mathcal{E}_{v}^{(t)} = \psi_j$ refers to energy consumption associated with the deployment of the $j^{th}$ VNF instance where $\psi_j$ is a constant value. The energy consumed by processor $i$ in Watts is $\mathcal{E}_{p}^{(t)} = \sigma^{*} P_{i}^{3}$ where $\sigma^{*}$ is a parameter determined by the processor structure. The wireless transmission power for UE $m$ is given by $\mathcal{E}_{w}^{(t)} = \frac{1}{\rho}||W_m||_{2}^{2}$ where $W_m$ is the precoding vector from all cells to UE $m$, and $\rho$ denotes the efficiency of power amplifier \cite{FOG} at the cells. Finally we have:
\begin{equation}
   \mathcal{E}_{Net}^{(t)} =  \sum_{i=1}^{I}\sigma^{*} P_{i}^{3}+\sum_{j=1}^{J} \psi_{j}  +  \sum_{m=1}^{M}\frac{1}{\rho}||W_m||_{2}^{2}
\end{equation}
We define overall network cost as all costs incurred at each time step:
\begin{equation}
   \mathcal{N}_{T}^{(t)} = \frac{\omega_1^{*} {K}_{Net}^{(t)} + \omega_2^{*} \mathcal{L}_{Net}^{(t)} + \omega_3^{*}\mathcal{E}_{Net}^{(t)}}{M}  
\end{equation}
Where $\omega_1^{*},\omega_2^{*},\omega_3^{*} \in \mathbb{R}$ are fixed weights which can be set based on the operator preferences.
\section{MDP and building NS environment by Gym}
In this section, we formulate resource allocation optimization problem as Markov Decision Process (MDP). We contemplate the autonomous CU with the goal of improve average return. To this end, we define the observation space and action space that CU can take at each time step.

The MDP for a single agent often is defined by a 5-tuple $(S,A,P,\gamma,R)$. consisting of a set of states $S$ (state space), a set of actions $A$ (action space), $P$ denotes the state transition probability for state $s$ and action $a$. The key term of MDP is decision. In fact, the way that agent make decisions for what actions to do in what states is called a policy which denotes with the symbol $\pi$. The notation of return ${G}_t$ refers to total discounted rewards from time step $t$ and the main goal is to maximize this return, ${G}_t = \sum_{n=0}^{\infty}\left(\gamma^{n} R_{t+n+1}\right)
$. Where $\gamma$ is a real-valued discount factor weighting of future rewards and $\gamma \in [0,1]$ refers to how much we value rewards right now relative to rewards in the future as short-sighted $(\gamma=0)$ or far-sighted $(\gamma=1)$.

The value function informs the agent how good is at each state or action and how much reward to expect takes a particular $V({s}_t,{a}_t)$.
Another value function $Q$ which not only depends on the state $s$  but also action $a$ is action-value function. The policies determine relation between $Q$ and $V$. The optimal policy in Reinforcement Learning (RL) is the best policy for which there is no greater value function, so for optimal value functions and optimal action-value function we have $\forall{s \in S,}\hspace{5mm}{V}_{\star}(s)={\max}_{\pi}\{{V}_{\pi}(s)\}$ and $\forall{s \in S,}\hspace{2mm}{a \in A,}\hspace{5mm}{Q}_{\star}(s,a)={\max}_{\pi}\{{Q}_{\pi}(s,a)\}$ respectively.

A continuous state and action space in  OpenAI Gym is defined the action that an agent can take and the input that the agent receives are both continuous values:

\begin{enumerate}
  \item \textbf{State Space:} We use Box space as multidimensional continuous spaces with bounds. In telecom environment the state space is the set of possible network configurations. We consider state at time step $t$ consists of :
  \begin{itemize}
  \item The number of new UEs which connect to network and request services for each slice $(X^{(t)})$.
  \item Computing resources allocated to each VNF $ (C^{(t)})$
  \item Delay status with respect to latency cost for each slice $( \mathcal{L}^{(t)})$
  \item Energy status with respect to energy cost for each slice$( \mathcal{E}^{(t)} )$
  \item Number of users being served in each slice $(m^{(t)})$ 
  \item Number of VNF instantiations in each slice $ (V^{(t)})$
  \end{itemize}
  
  The network state space or input can be characterized by $S^{(t)} = \{X^{(t)}, C^{(t)},  \mathcal{L}^{(t)},  \mathcal{E}^{(t)}, m^{(t)},  V^{(t)}\}$.
      \item \textbf{Action Space:} We consider vertical scaling action space. The vertical scaling can be classified into scale up and scale down that is related to  increasing or decreasing capacity respectively. The CU select continuous value action with respect to traffic fluctuation and learn to decide to increase/decrease computing resources allocated to each VNF. In OpenAI Gym, It takes an action as input and provides observation, reward, done and an optional info object as output at each step. Let consider vertical scaling action for CPU resources as $\zeta_{CPU}^{(t)}$. Therefore, due to change the allocation resources according to time slot, we have:
      \begin{equation}
       \zeta_{CPU}^{(t)} \in \{ z  |   z \in \mathbb{R},-K_{Net}^{(t)}\leq z \leq K_{I}^{(t)}   -K_{Net}^{(t)}\}
      \end{equation}
      One may note that vertical scaling is limited by the amount of free computing resources available on the physical server hosting the virtual machine \cite{SAAS}. 
  \item \textbf{Reward:} The main objective of this work is minimize the total network cost where the agent learns to increase the expected return. To this end we define the return as follows:
  \begin{equation}
    R^{(t)} = \frac{1}{ \mathcal{N}_{T}^{(t)}}
  \end{equation}
  We pursue an experimental approach because maybe the total network cost ($ \mathcal{N}_{T}^{(t)}$) is a general and imprecise  metric to guide the agent for learning and leading to good results. Consequently, tuning the hyperparameters, Deep Neural Networks (DNNs) architecture and designing training steps are very tricky.
\end{enumerate}

\section{Twin Delayed DDPG}
 The basic idea behind policy-based algorithms is to adjust the parameters $\phi$ of the policy in the direction of the performance gradient $\nabla_{\phi}J(\pi_{\phi})$. The fundamental result underlying these algorithms is the policy gradient theorem \cite{DDPG}: $\nabla_{\phi}J(\pi_{\phi})=\int_{S}p_{\pi}(s)\int_{A}\nabla_{\phi}\pi_{\phi}(a|s)Q^{\pi}(s,a)dads.$

We can parameterize policy like value function and the goal is to find the optimal policy $\pi_{\phi}$ where $\phi$ includes updating the weight of the policy. The expected return can be approximated in many ways. We calculate the gradient of expected return according to parameters of $\phi$ as $\nabla_{\phi}J(\phi)$. We use gradient ascent as opposed of gradient descent for updating the parameters, ${\phi}_{t+1}={\phi}_{t}+\alpha\nabla_{\phi}J({\pi}_{\phi}) | {\phi}_{t}$.
\begin{figure}[ht!]
\centering
\includegraphics[scale=0.27]{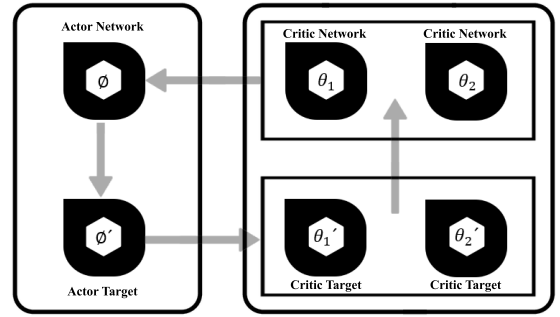}
\caption{Training flow process between different DNNs.}

\end{figure}

\begin{algorithm}[ht!]
\caption{TD3-based NS with OpenAI Gym}
\scriptsize
\SetAlgoLined
 Initialize actor network $\phi$ and critic networks $\theta_1$, $\theta_2$
 
 Initialize (copy parameters) target networks ${\phi}^{\prime}$,  ${\theta}_1^{\prime}$, ${\theta}_2^{\prime}$
 
 Initialize replay buffer $\beta$
 
 Import network slicing environment (`smartech--v0')
 
 \While {t < max\_timesteps}{

  \eIf{t < start\_timesteps}{
   $a$ = env.action\_space.sample()
   }{
   $a \longleftarrow {\pi}_{\phi}({s})+\epsilon,  \;\;\; \epsilon \sim \mathcal{N}(0,{\sigma})$
   
  }
  next\_state, reward, done, \_ = env.step($a$)
  
  store the new transition ${({s}_t,{a}_t,{r}_t,{s}_{t+1})}$ into $\beta$
  
   \If{t $\geq$ start\_timesteps}{
   sample batch of transitions ${({s}_{t_{B}},{a}_{t_{B}},{r}_{t_{B}},{s}_{t_{B}+1})}$
  
   $\tilde{a} \longleftarrow {\pi}_{\phi}{\prime}({s}^{\prime})+\epsilon,  \;\;\; \epsilon \sim clip(\mathcal{N}(0,\tilde{\sigma}), -c, c)$
   
   ${Q}_t = r+\gamma * \min({Q}_{t1}^{\prime},{Q}_{t2}^{\prime})$
   
   $L = {l}_{MSE}({Q}_1,{Q}_t)+{l}_{MSE}({Q}_2,{Q}_t)$
   
   ${\theta}_{f} \longleftarrow argmin_{\theta_f} {N}^{-1}\sum(L- Q_{\theta_f(s,a)})^{2}$
   
   \If{$t \% policy\_freq == 0$}{ $\nabla_{\phi}J(\phi)={N}^{-1}\sum\left[\nabla_{a}{Q}_{\theta_1}(s,a) |_{a={\pi(\phi)}}\nabla_{\phi}\pi_{\phi}(s)\right]$
   
${\theta}_{f}^{\prime} \longleftarrow \tau {\theta}_{f} + (1 - \tau){\theta}_{f}^{\prime}$

${\phi}^{\prime} \longleftarrow \tau {\phi} + (1 - \tau){\phi}^{\prime}$

}

   }
  
  \If{done}{
   obs, done = env.reset(), False
   }
  t=t+1
 }
\end{algorithm}

In actor-Critic method, we have two models that work concurrently where the actor is a policy taking state as input and delivering actions as output, while the critic takes states and actions concatenated together and return the Q-value and a policy that can be updated through the deterministic policy gradient\cite{TD3}, $\nabla_{\phi}J(\phi)=\mathbb{E}_{s\sim p_{\pi}} \left[\nabla_{a}{Q}^{\pi}(s,a) |_{a={\pi(s)}}\nabla_{\phi}\pi_{\phi}(s)\right]$. where ${Q}^{\pi}(s,a) = \mathbb{E}_{{s}_f\sim p_{\pi}\sim {a}_f\sim \pi}\left[{R}_t | s,a\right] $ is known as value function or critic. 

Initially we should store random experience in the buffer $\beta$. In the other words, we store ${({s}_t,{a}_t,{r}_t,{s}_{t+1})}$ to train Deep Q-Network. We take a random batch $B$ and for all transitions ${({s}_{t_{B}},{a}_{t_{B}},{r}_{t_{B}},{s}_{t_{B}+1})}$ of $\beta$, the predictions are $Q({s}_{t_{B}},{a}_{t_{B}})$ and the targets consider as optimal immediate return that are exactly first part of temporal difference learning (TD) error as $R({s}_{t_{B}},{a}_{t_{B}})+\gamma{\max}_{a}({Q}_{({s}_{t_{B}+1},a)})$, and over the whole batch $B$, we calculate the loss between predictions and the targets in the batch $B$. Another target network is used instead of using Q-network to calculate the target to fulfill more stability for learning algorithm. As shown in figure 2, the TD3 is based on the actor-critic model that it leverages three tricks to improve algorithm:

\begin{enumerate}
  \item \textbf{Clipped double Q-learning with pair of critic networks:} 
  We use two DNNs as two actor networks and denote them by $\phi$ as actor network and ${\phi}^{\prime}$ as actor target. In addition, we create two pair of critic networks and denote them by ${\theta}_1$,${\theta}_2$ for parameterization of value network and ${\theta}_1^{\prime}$,${\theta}_2^{\prime}$ as critic targets. Indeed, two learnings happen simultaneously, namely, Q-learning and Policy learning, and they address approximation error, reduce the bias, and find the highest Q-value. This was inspired by the technique seen in \cite{Double} as Double-Q Learning. For each element and transition of batch, the actor target plays ${a}^{\prime}$ based on ${s}^{\prime}$ while we add Gaussian noise to this ${a}^{\prime}$. The critic targets takes the couple (${s}^{\prime},{a}^{\prime}$) and return two Q-values  ${Q}_{t1}^{\prime}$ and ${Q}_{t2}^{\prime}$ as output. Then, the ($\min {Q}_{t1}^{\prime}$,${Q}_{t2}^{\prime}$) is considered as an approximated value for critic networks. In \cite{DLDQ} has proposed using the target network as one of the value estimates. Given that we calculate the final target of the two value networks, we have:
\begin{equation}
  {Q}_t = r+\gamma * \min({Q}_{t1}^{\prime},{Q}_{t2}^{\prime})
\end{equation}
  then the two critic networks return two Q-values as ${Q}_1$(s,a) and ${Q}_2$(s,a). Next, we calculate the loss based on two critic networks and with Mean Squared Error (MSE). To minimize the loss over iterations via back-propagation technique, we use an efficient optimizer called Adaptive  Moment Estimation  (Adam) \cite{ADAM} in our code:
\begin{subequations}
\begin{equation}
L = {l}_{MSE}({Q}_1,{Q}_t)+{l}_{MSE}({Q}_2,{Q}_t)
\end{equation}
\begin{equation}
\nabla_{\phi}J(\phi)={N}^{-1}\sum\left[\nabla_{a}{Q}_{\theta_1}(s,a) |_{a={\pi(\phi)}}\nabla_{\phi}\pi_{\phi}(s)\right]
\end{equation}
\end{subequations}
  In the next step, we explain how we update the target networks.
  \item \textbf{Delayed policy updates and target networks:}
  The main idea is to update the policy network less frequently than the value network since we need to estimate the value with lower variance\cite{CCDR}. The update rule is given by Polyak Averaging, so we update parameters by:
  \begin{subequations}
  \begin{equation}
  {\theta}_{f}^{\prime} \longleftarrow \tau {\theta}_{f} + (1 - \tau){\theta}_{f}^{\prime}
  \end{equation}
  \begin{equation}
  {\phi}^{\prime} \longleftarrow \tau {\phi} + (1 - \tau){\phi}^{\prime}
  \end{equation}
  \end{subequations}
  where $\tau \leq1$ is an hyperparameter to tune the speed of updating.

  \item \textbf{Target policy smoothing and noise regularisation:} 
 When updating the critic, a learning target using a deterministic policy is highly susceptible to inaccuracies induced by function approximation error, increasing the variance of the target. This induced variance can be reduced through regularization \cite{TD3} to be sure for the exploration of all possible continuous parameters. We add Gaussian noise to the next action ${a}^{\prime}$ to prevent two large actions played and disturb the state of the environment:
\begin{equation}
\tilde{a} \longleftarrow {\pi}_{\phi}{\prime}({s}^{\prime})+\epsilon,  \;\;\; \epsilon \sim clip(\mathcal{N}(0,\tilde{\sigma}), -c, c)
\end{equation}
  where the noise $\epsilon$ is sampled from a Gaussian distribution with zero and certain standard deviation and clipped in a certain range of value between $-c$ and c to encourage exploration. Due to avoid the error of using the impossible value of actions, we clip the added noise to the range of possible actions (min\_action, max\_action). The TD3-based NS method is summarized in Algorithm 1.
\end{enumerate}

\section{Numerical Results}
To evaluate our method described in section \rom{4}, we generate six deep neural networks which work together based on actor-critic model.
\begin{figure}[ht!]
\centering
\includegraphics[scale=.6]{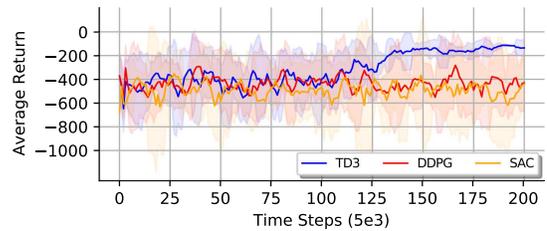}
\caption{Learning curves for the gym NS environment.}
\end{figure}
The implementation is written in PyTorch, followed by the experimental parameters used in the simulations.
\begin{figure*}[ht!]
\centering
\subfloat[Admission rate - Slice 1]{%
      \includegraphics[width=0.24\textwidth]{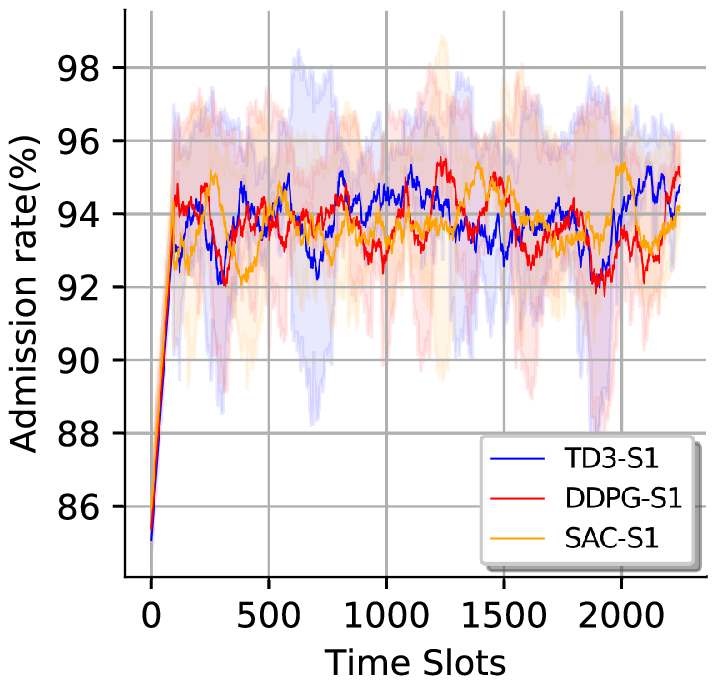}}
      \hfill
\subfloat[Admission rate - Slice 2]{%
      \includegraphics[width=0.24\textwidth]{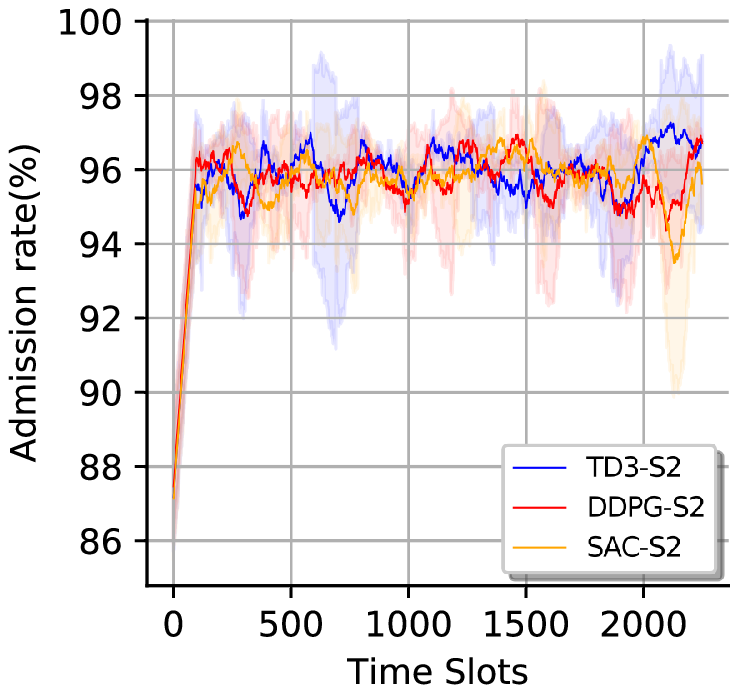}}
      \hfill
\subfloat[Latency - Slice 1]{%
      \includegraphics[width=0.24\textwidth]{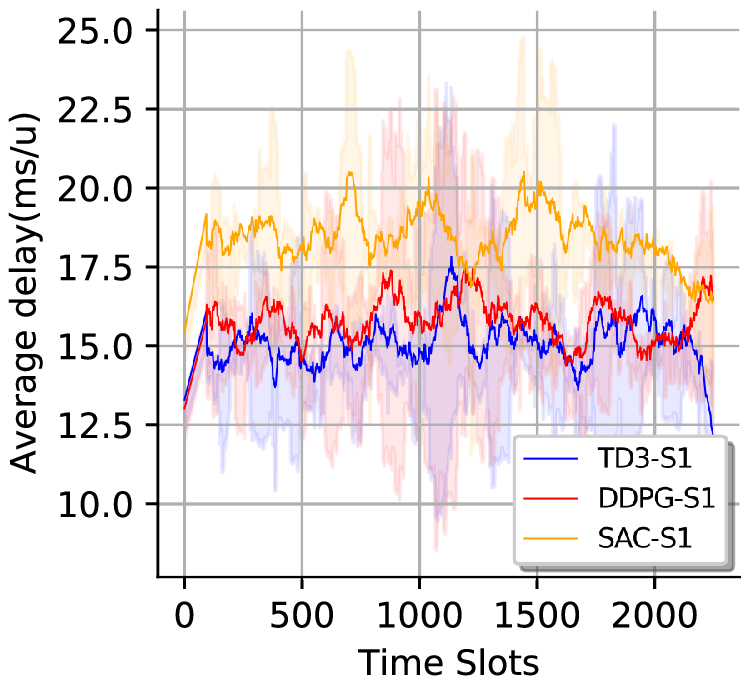}}
      \hfill
\subfloat[Latency - Slice 2]{%
      \includegraphics[width=0.24\textwidth]{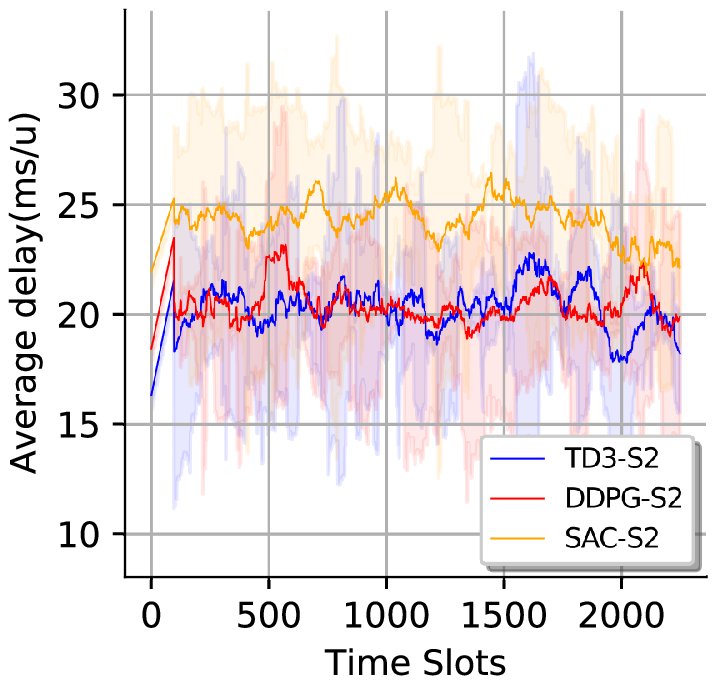}}\\
\subfloat[Dealy QoS violation - Slice 1]{%
      \includegraphics[width=0.24\textwidth]{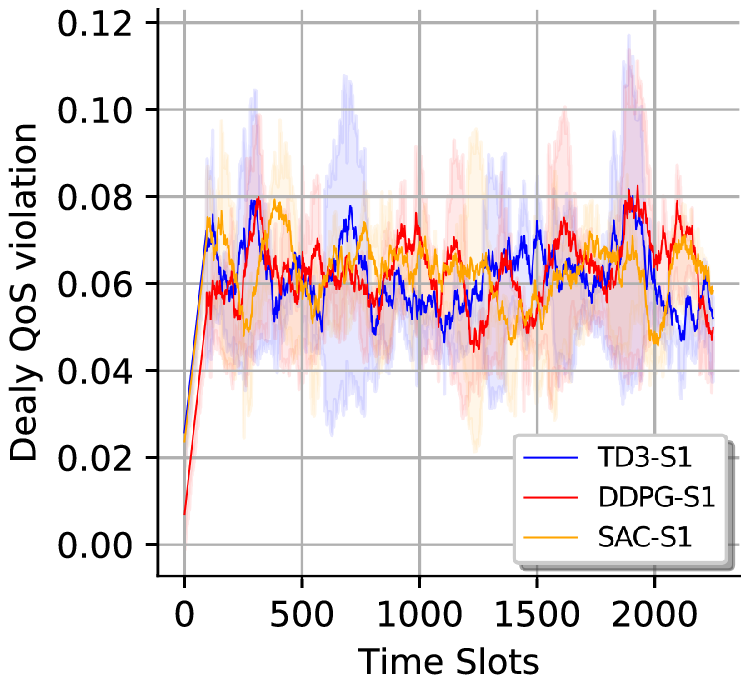}}
      \hfill
\subfloat[Dealy QoS violation - Slice 2]{%
      \includegraphics[width=0.24\textwidth]{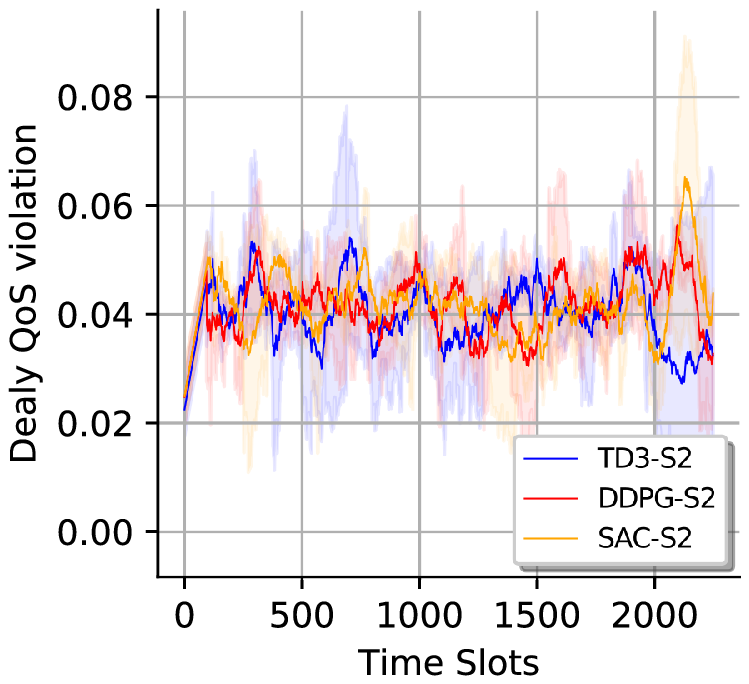}}
      \hfill
\subfloat[Energy consumption - Slice 1]{%
      \includegraphics[width=0.24\textwidth]{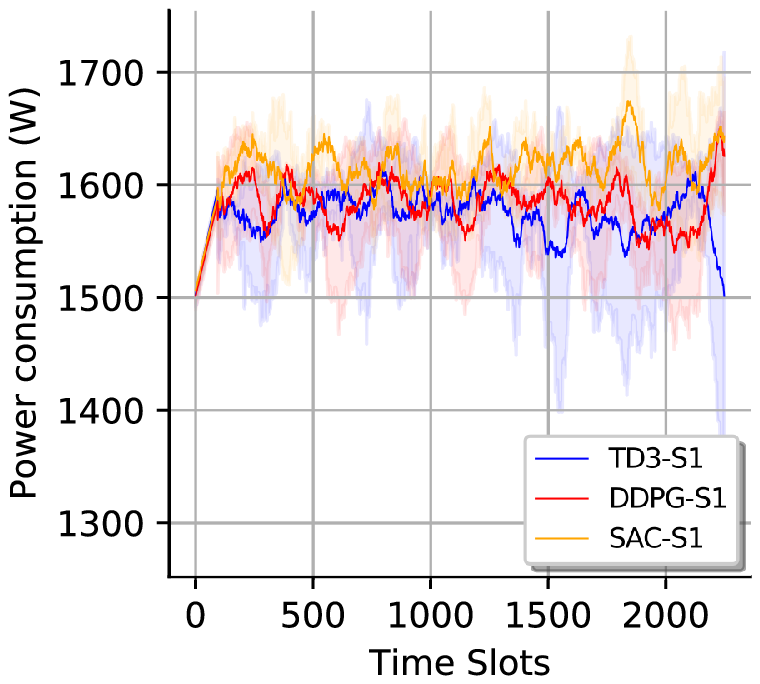}}
      \hfill
\subfloat[Energy consumption - Slice 2]{%
      \includegraphics[width=0.24\textwidth]{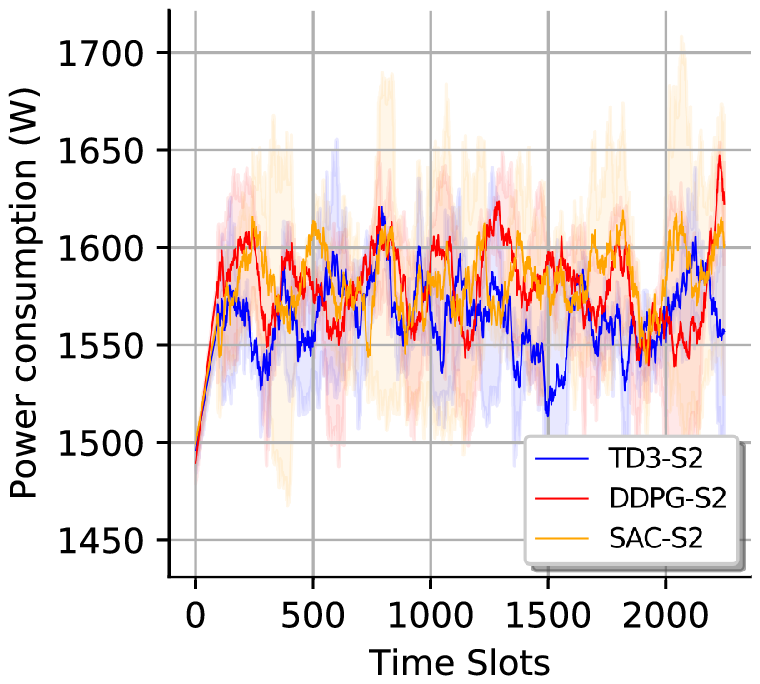}}\\
\subfloat[CPU utilization - Slice 1]{%
      \includegraphics[width=0.24\textwidth]{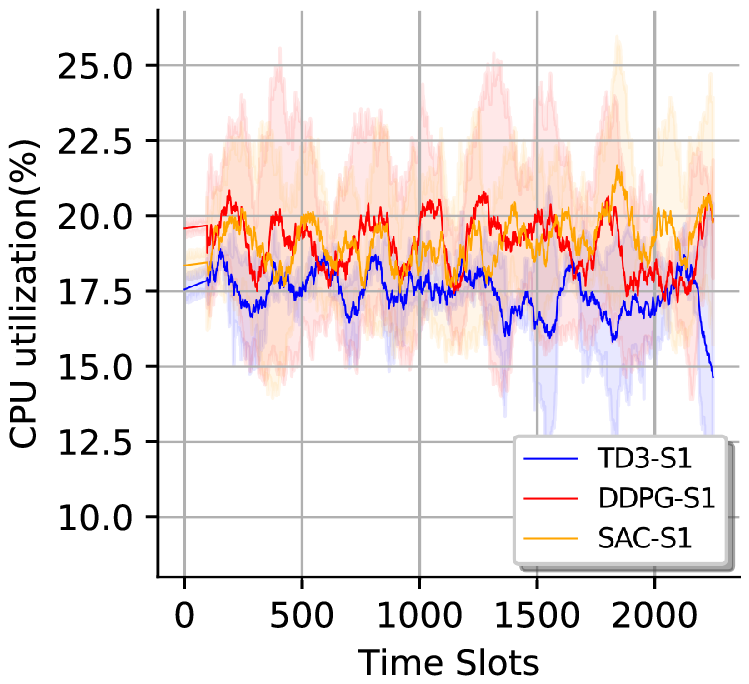}}
      \hfill
\subfloat[CPU utilization - Slice 2]{%
      \includegraphics[width=0.24\textwidth]{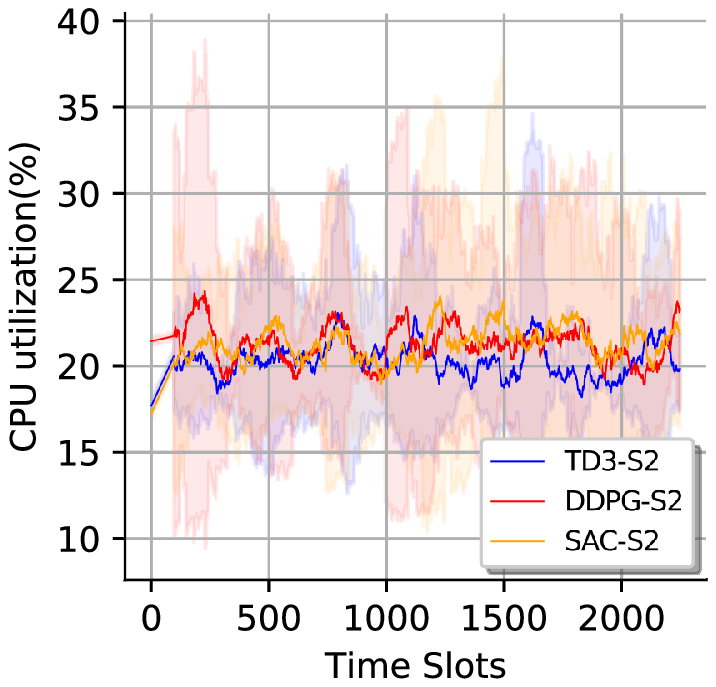}}
      \hfill
\subfloat[Network latency]{%
      \includegraphics[width=0.24\textwidth]{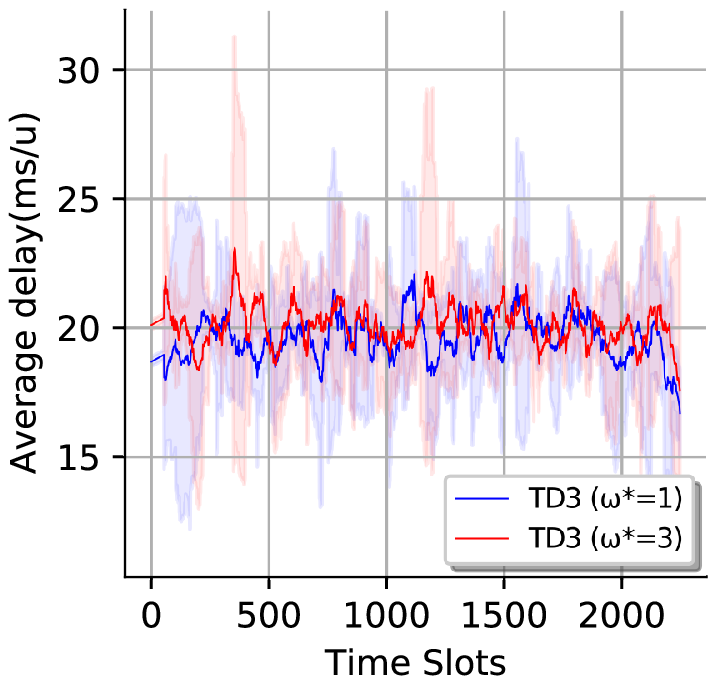}}
      \hfill
\subfloat[Network energy consumption]{%
      \includegraphics[width=0.24\textwidth]{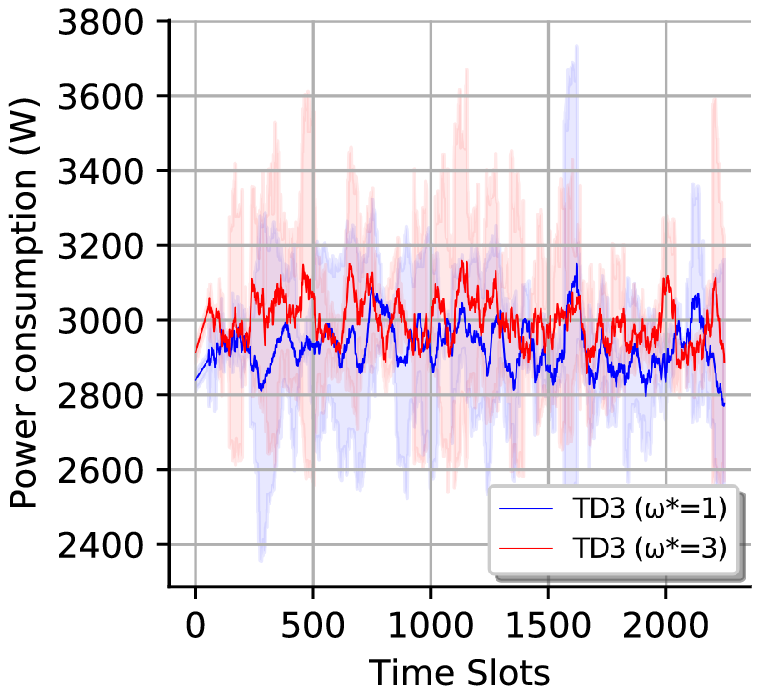}}
\caption{Network performance and costs comparison between TD3 and other DRL benchmarks. The curves are smoothed for visual clarity with respect to confidence bands and standard deviation. [\footnotesize$\beta$\_Initialization = $20000$, B\_size = $128$, Policy\_freq = $2$, Policy\_noise = $0.2$, Noise\_clip = $0.5$, Exploration\_noise = $0.1$, $\tau$ = $0.005$, Optimizer : $Adam$ and $SGD$, Activation\_function : $ReLU$ and $tanh$, Discount\_factor = $0.99$, N  = $10$, QoS\_S1 requirement $\eta_{m}^{(t)}$  = $20 (ms)$, QoS\_S2 requirement $\eta_{m}^{(t)}$  = $40 (ms)$, $\sigma^{*}$ = $10^{-26}$, $P$ = $10^{9}$]}

\end{figure*}
In fact, the values of parameters depend highly on capability, scenario and technology used. We measure the performance on our customized NS environment, interfaced through OpenAI Gym to fulfill reproducible comparison. In this environment, the mobile network operator (MNO) collects the free and unused resources from the tenants and when slices need more resources can receive new resources.It is done either periodically to avoid over-heading or based on requests of tenants. We consider a two-tenants scenario, i.e., two slices with different QoS requirements in terms of latency and CPU constraints. For each time step, users packets arrive into the network and the algorithm computes the computing requirements to allocate to the relevant VNF. We compare the performance of TD3 method against a fine-tuned version of the DDPG presented in \cite{CCDR} and \cite{TD3} as well as Soft Actor-Critic (SAC) \cite{SAC} to keep all algorithms consistent. Table I presents the number of DNNs for each method. As shown in Figure 3, the learning curve of TD3 outperforms all other algorithms in final performance with respect to our big and complex state space. However, the problem formulation is general, we use constraints as penalty in implementation to leads agent to the good results and this is the reason of negative values in the learning curves.
\begin{table}[ht!]
\centering
\caption{Number of networks for algorithms.}
\begin{tabular}[t]{lccc}
\hline
&SAC    &DDPG    &TD3\\
\hline
Policy&1&1&1\\
Value&2&1&2\\
Target Policy&1&1&1\\
Target Value&2&1&2\\
\hline
\end{tabular}

\end{table}

In Figure 4, we present the results of comparison with respect to network performance and cost functions in (\rom{2}) for each slice under similar traffic patterns and also network measures based on different weights that are shown in Figures 4-(k) and 4-(l).\\
\textbf{Admission rate:} Figures 4-(a) and 4-(b) show that the re-tuned TD3 algorithm outperforms the other approaches with respect to resource availability and constraints. The algorithm learns according to iterations or interact with the environment with different configurations of network so this is the reason of high fluctuation. The main goal for the algorithm is to find the best policy during the time. The similarity between results of slice 1 and slice 2 is because the MNO is trading-off between slices while there is a high resource availability. This approach enables slices to request more resources and results in provisioning services to more users and increasing admission rate.\\
\textbf{Latency:} As shown in figure 4-(c) and 4-(d), our solution leads to less average delay per user compared to DDPG and SAC.\\
\textbf{Dealy QoS violation:} The comparison between Delay QoS metric of TD3 and other schemes are presented in figure 4-(e) and 4-(f).\\
\textbf{Energy consumption:} Figures 4-(g) and 4-(h) show that the performance of our scheme and other methods where agent learns to satisfy another objective and minimize power consumption by decreasing VNFs instantiation and tuning wireless transmission power.\\
\textbf{CPU utilization:} As depicted in figures 4-(i) and 4-(j), the TD3 deployment leads to more efficient usage of CPU compared to other methods. 

\section{Conclusion}

In this paper, we have presented a new continuous multi-objective zero-touch NS solution. In this intent, we have developed an OpenAI Gym NS environment and used advanced DRL-based algorithm for resource allocation problem to enable CU to learn how to re-configure computing resources autonomously, aiming at minimizing the latency, energy consumption and VNF instantiation for each slice. This method leverages 3 techniques to  fulfill more stability of learning algorithm. In this respect, we have compared the network performance and costs between TD3 and other DRL benchmarks. We have shown that the proposed solution outperforms other DRL methods.

\section*{Acknowledgement}This work has been supported in part by the research projects 5G STEP FWD (722429), MonB5G (871780),   5G-SOLUTIONS (856691), AGAUR(2017-SGR-891) and SPOT5G (TEC2017-87456-P).

\end{document}